\documentclass[12pt]{article}
\usepackage{graphicx}
\usepackage{amssymb}

\begin{document}

\setlength{\baselineskip}{18pt}
\begin{titlepage}
\begin{flushright}
Alberta Thy  20-06 
\end{flushright}

\centerline{{\large \bf   Coulomb force correction to the decay $b \rightarrow c\bar{c} s$  in the threshold }}

\vspace{8mm}

\centerline{ K. Hasegawa
\footnote{hasegawa@phys.ualberta.ca}}
\vspace{5mm}
\centerline{{\it Department of Physics, University of Alberta, Edmonton, AB T6G 2J1, Canada  }}

\vspace{15mm}
\centerline{\large Abstract}
\vspace{2mm}
We study the physical origins of the ${\cal O}(\alpha_{s})$ and ${\cal O}(\alpha_{s}^{2})$ 
corrections to the $\bar{c}-s$ current in the decay $b \rightarrow c\bar{c} s$ in the threshold region
$\delta=(M_{b}-2m_{c})/2M_{b} \ll 1$. We obtain the corrections which are produced by the Coulomb
force between the anti-charm and strange quarks. The Coulomb corrections $C_{F}\pi^{2}$ at 
${\cal O}(\alpha_{s})$ and $-C_{F}^{2}\pi^{2}\mbox{ln} \ \delta$ at ${\cal O}(\alpha_{s}^{2})$ 
account for 300\% and 120\% of the corresponding terms in the Abelian-type perturbative corrections
respectively. The differences between the Coulomb and perturbative corrections imply that 
other corrections which have other physical origins exist.
We also check that the Wilson coefficient for the anomalous dimension of the $\bar{c}-s$
current reproduces the leading and the next-to-leading logarithmic terms in the perturbative corrections.

\end{titlepage}

\section{Introduction} 

Disagreement between the experimental measurements and the theoretical predictions for the
branching ratio of the semi-leptonic decay of the B meson, $\mbox{BR}(B \to X \bar{l} \nu_{l})$, 
has been reported \cite{big}. Because the non-perturbative corrections related to the confinement of the 
participating quarks inside the hadrons are at a level of a few percent \cite{big}, the branching ratio of 
the semi-leptonic decay is mainly determined by the decay rates of the three decay modes of the
free $b$ quark decay, the semi-leptonic mode $b \rightarrow c \bar{l} \nu_{l}$, the non-leptonic 
mode with light quarks $b \rightarrow c \bar{u} d$, and the non-leptonic mode with an extra 
charm quark $b \rightarrow c \bar{c} s$. 
Here, the non-leptonic modes contribute to the branching ratio of the semi-leptonic decay
through the total decay width of the $b$ quark.
About 30\% enhancement of the ${\cal O}(\alpha_{s})$ corrections to the $\bar{c}-s$ current 
in the non-leptonic mode $b \rightarrow c \bar{c} s$ has been obtained \cite{hok,bag, bag2, bag3, vol1}.
When the large enhancement  at the ${\cal O}(\alpha_{s})$ level and 
the evolution of the renormalization group  from the scale of the $W$ boson mass to that of the $b$ quark mass are taken
into account,  the significant disagreement with experiments vanishes \cite{bag, bag2, bag3,neu}. 
It has  also been reported that the ${\cal O}(\alpha_{s}^{2})$ corrections to the non-leptonic 
mode  $b \rightarrow c \bar{u} d$ remain in agreement with experiments \cite{cza1}.
At the present stage, the ${\cal O}(\alpha_{s}^{2})$ corrections to the other non-leptonic mode, 
$b \rightarrow c \bar{c} s$, becomes one of the main issues with the theoretical predictions of the 
semi-leptonic branching ratio. The ${\cal O}(\alpha_{s})$ correction to the $\bar{c}-s$ current in the
mode $b \rightarrow c \bar{c} s$ is a monotonically increasing function of the mass of the charm quark, and
the ratio of the ${\cal O}(\alpha_{s})$ correction to the tree-level rate diverges in the threshold region 
$\delta=(M_{b}-2m_{c})/2M_{b}\ll 1$ due to the logarithmic term $(\mbox{ln} \ \delta)$. 
Furthermore, although the fully analytic form of the ${\cal O}(\alpha_{s}^{2})$ corrections,
including the charm mass dependence, has not been obtained, it is known that the ${\cal O}(\alpha_{s}^{2})$ 
corrections also contain the leading logarithmic terms $(\mbox{ln} \ \delta)^{2}$ and the next-to-leading 
logarithms $(\mbox{ln} \ \delta)$. We conjecture that the large corrections come from the threshold region
and think it meaningful to elucidate the physical origins of the large corrections. With this goal in mind, in the present 
paper, we focus on the ${\cal O}(\alpha_{s})$ and ${\cal O}(\alpha_{s}^{2})$ corrections to the $\bar{c}-s$ 
current in the threshold region of the decay $b \rightarrow c\bar{c} s$ and estimate the quantum 
corrections which come from the two physical origins, the Coulomb force between the $\bar{c}$ and $s$
quarks, and the anomalous dimension of the $\bar{c}-s$ current.

It is well known that the Coulomb force between the proton and the electron produces the additional
corrections to the tree-level decay rate of the neutron beta decay $n \to p e^{-} \nu_{e}$ (see, e.g, 
the textbook \cite{kon}). The wave function of the electron is distorted by the Coulomb force
generated by the electromagnetic charge of the approximately static proton. The distorted wave function 
forms the Fermi function in the decay rate. Considering the Fermi function in the neutron beta decay, 
here we attempt to incorporate the effects of the Coulomb-like gluon exchanges between the $\bar{c}$ and $s$
quarks in the threshold region of the decay $b \rightarrow c\bar{c} s$. In Ref.\cite{vol2} , it is pointed out that
the Fermi function for the $\bar{c}-s$ current produces a next-to-leading logarithm in the ${\cal O}(\alpha_{s}^{2})$
corrections. However, in that work, the Fermi function  is not introduced to the decay rate formula
which contains the phase space integrals of the final states.
We believe it is worthwhile to obtain an estimate of the effect 
of the Coulomb distortion of the strange quark in the decay rate formula in order to obtain the more precise
predictions of the Coulomb corrections. Considering this point, the main purpose of the present paper is to 
incorporate the Fermi
function into the decay $b \rightarrow c\bar{c} s$ and to obtain the corrections which are produced by the
Coulomb force between the $\bar{c}$ and $s$ quarks in the threshold region. We then compare the 
Coulomb corrections with the Abelian parts of the perturbative corrections at the ${\cal O}(\alpha_{s})$ and
${\cal O}(\alpha_{s}^{2})$ levels. The other origin of the logarithmic terms is known. It is shown that
the Wilson coefficients for the anomalous dimension of the heavy-light current, which is called the hybrid
anomalous dimension, reproduce the leading logarithmic terms in the perturbative corrections of the two-body
 decay into one heavy and one light particle \cite{shi, pol1, pol2}. Furthermore, when the Wilson 
coefficients are improved by the non-logarithmic terms at ${\cal O}(\alpha_{s})$ in the perturbative 
corrections, the improved coefficient can produce the next-to-leading logarithmic terms in the 
${\cal O}(\alpha_{s}^{2})$ corrections \cite{jix, bro, bag4, che}. In the present paper, we confirm that the
leading and next-to-leading logarithms in the perturbative corrections to the decay rate 
$\Gamma(b \rightarrow c\bar{c} s)$ can also be reduced to the Wilson coefficients for the hybrid anomalous
dimension.

The remainder of the present paper is organized as follows. In \S \ref{two} we obtain the threshold expansion
of the perturbative corrections at ${\cal O}(\alpha_{s})$ and ${\cal O}(\alpha_{s}^{2})$ in the decay 
$b \rightarrow c\bar{c} s$.  In \S \ref{three} we obtain the Coulomb force corrections and compare
them with the Abelian parts of the perturbative corrections. In \S \ref{four} we show that the
Wilson coefficient for the hybrid anomalous dimension reproduces the leading and next-to-leading
logarithmic terms in the perturbative corrections. In \S \ref{five} we give a summary.

\section{Results of the perturbative calculation  \label{two}}
We start with the effective Lagrangian for the decay $b \rightarrow c\bar{c} s$, 
\footnote{The effective Lagrangian should include the Wilson coefficient which comes from 
the evolution of the renormalization group from the scale of the $W$ boson mass to that of the bottom quark mass
\cite{bag, bag2, bag3,neu}. But in order to compare the perturbative corrections with the Coulomb 
corrections, which is done in \S \ref{three}, we do not need the Wilson coefficient, and we therefore omit it for simplicity
 hereafter.}
\begin{eqnarray}
{\cal L}= -V_{cb}V_{cs}^{*}\frac{G_{F}}{\sqrt{2}} \bigr[ \bar{\psi_{c}} \gamma^{\mu}
(1-\gamma_{5})\psi_{b} \bigl] \bigr[ \bar{\psi_{s}} \gamma_{\mu}(1-\gamma_{5})\psi_{c} \bigl]. \label{lag}
\end{eqnarray}
We first calculate the tree-level decay rate by using the ordinal decay rate formula
\begin{eqnarray}
 \Gamma (b \rightarrow c \bar{c} s) = \frac{1}{2M_{b}} 
 \int 
\frac{d^{3}p_{c}}{(2\pi)^{3}} \frac{1}{2 p_{c}^{0}} 
\frac{d^{3}p_{\bar{c}}}{(2\pi)^{3}} \frac{1}{2 p_{\bar{c}}^{0}} 
\frac{d^{3}p_{s}}{(2\pi)^{3}} \frac{1}{2 p_{s}^{0}} 
|{\cal M}|^{2}   \nonumber \\
(2\pi)^{4} \delta^{(4)}(p_{b}-p_{c}-p_{\bar{c}}-p_{s}),
\label{direct}
\end{eqnarray}
where ${\cal M}$ is the invariant matrix and $M_{b}$ is the mass of the bottom quark.
The momenta of the bottom, charm, anti-charm, and strange quarks are written $p_{b}, p_{c}, p_{\bar{c}}$,
\mbox{and} $p_{s}$. The tree-level decay rate $\Gamma (b \rightarrow c \bar{c} s)_{0}$ is obtained as
\begin{eqnarray}
\Gamma (b \rightarrow c \bar{c} s)_{0}
&=& \Gamma_{\mu}\Biggl[
\sqrt{1-4\epsilon^{2}}\bigl( 1-14\epsilon^{2} -2\epsilon^{4}-12\epsilon^{6} \bigr)
\nonumber \\
&&\hspace{30mm} +24\epsilon^{4}(1-\epsilon^{4}) \mbox{ln} \Biggl(\frac{1+\sqrt{1-4\epsilon^{2}}}
{1-\sqrt{1-4\epsilon^{2}}}\Biggr)
\Biggr],  \label{treeres}
\end{eqnarray}
where the ratio of the charm quark mass to the bottom quark mass is written $\epsilon=m_{c}/M_{b}$. The decay rate
at $\epsilon=0$ is denoted by $\Gamma_{\mu}=N_{c}|V_{cb}|^{2}|V_{cs}|^{2}  G_{F}^{2} M_{b}^{5}/(192 \pi^{3})$,
with $N_{c}=3$. We can also obtain the decay rate from the so-called `factorization formula',
\begin{equation}
\Gamma (b \rightarrow c \bar{c} s ) =
|V_{cb}|^{2}|V_{cs}|^{2} G_{F}^{2} \int^{(M_{b}-m_{c})^{2}}_{m_{c}^{2}} \frac{dq^{2}}{(2 \pi)} 
\ \sqrt{q^{2}}  \  X^{\mu \nu} Y_{\mu \nu}.   \label{fact}
\end{equation}
Here, $X^{\mu \nu}$ and $Y_{\mu \nu}$ are defined as
\begin{eqnarray}
\Gamma(b \rightarrow c W^{-}) &=& \epsilon_{\mu}^{\lambda *} (q) 
\epsilon_{\nu}^{\lambda} (q) X^{\mu \nu}, \label{facx} \\ 
\Gamma (W^{-} \rightarrow \bar{c} s) &=& \epsilon_{\mu}^{\lambda} (q)^{*}
\epsilon_{\nu}^{\lambda} (q)  Y^{\mu \nu},  \label{facy}
\end{eqnarray}
where $q$ and $\epsilon_{\mu}^{\lambda} (q)$ are the momentum and polarization vector of the $W$ gauge boson. 
When we calculate $X^{\mu \nu}$ and $Y^{\mu \nu}$ in Eqs. (\ref{facx}) and (\ref{facy}), we use the Lagrangians
${\cal L}=\bar{\psi_{c}} \gamma^{\mu}(1-\gamma_{5})\psi_{b} W_{\mu}^{+}$ and ${\cal L}=\bar{\psi_{s}} 
\gamma^{\mu}(1-\gamma_{5})\psi_{c} W_{\mu}^{-}$, respectively, where the coupling constants are normalized
to unity. The mass squared of the $W$ boson, $q^{2}$, is the variable of integration in Eq. (\ref{fact}), and it runs over the 
region in which both of the decay processes $b \rightarrow c W^{-}$ and $W^{-} \rightarrow \bar{c} s$ can occur.
We can calculate $X^{\mu \nu}$ at tree level as
\begin{eqnarray}
 X^{\mu \nu} &=&
  \frac{1}{4\pi}\frac{\sqrt{\lambda(M_{b}^{2}, m_{c}^{2}, q^{2})}}{M_{b}^{3}}
 \int  \frac{d \Omega}{4\pi}  
   \nonumber  \\
& & \biggl[ 2p_{b}^{\mu} p_{b}^{\nu}- p_{b}^{\mu} q^{\nu}- p_{b}^{\nu} q^{\mu}
   -g^{\mu \nu} M_{b}^{2} +g^{\mu \nu} (p_{b} \cdot q)-i \epsilon^{\alpha \mu \beta \nu}
   q_{\alpha} p_{b, \beta}  \biggr],   \label{x}
\end{eqnarray}
where the variable $\Omega$ represents the direction of the momentum of the $W$ boson, and $\lambda$ is defined as
$\lambda(a,b,c)=a^{2}+b^{2}+c^{2}-2(ab+bc+ca)$. We can also calculate $Y^{\mu \nu}$ at tree level as
\begin{eqnarray}
Y^{\mu \nu}  = -\frac{N_{c}}{4\pi \sqrt{q^{2}}}\bigl[A_{0}(q^{2}g^{\mu \nu}-q^{\mu}q^{\nu})
 -B_{0}q^{\mu}q^{\nu}  \label{y}\bigr],
\end{eqnarray}
where $A_{0}$ and $B_{0}$ are the transversal and longitudinal parts, respectively, in the $W$ boson decay. They
are calculated as 
\begin{eqnarray}
A_{0}\biggl(\frac{\epsilon^{2}}{\omega^{2}}\biggr)&=&\frac{2}{3}
\biggl(1-\frac{\epsilon^{2}}{\omega^{2}} \biggr)^{2} 
\biggl(1+\frac{\epsilon^{2}}{2\omega^{2}} \biggr),  \label{a0} \\
B_{0}\biggl(\frac{\epsilon^{2}}{\omega^{2}}\biggr) &=& \frac{\epsilon^{2}}{\omega^{2}} 
\biggl(1-\frac{\epsilon^{2}}{\omega^{2}} \biggr)^{2}, \label{b0}
\end{eqnarray}
where $\omega = \sqrt{q^{2}}/M_{b}$. Substituting Eqs. (\ref{x}) and (\ref{y}) into Eq. (\ref{fact}), 
we obtain the following form :
\begin{eqnarray}
\Gamma (b \rightarrow c \bar{c} s )  &=&
 6 \Gamma_{\mu} \int_{\epsilon^{2}}^{(1-\epsilon)^{2}} d\omega^{2}   \sqrt{\lambda(1, \epsilon^{2}, \omega^{2})}
\nonumber  \\
 &&\biggl[\omega^{2}(1+\epsilon^{2}-\omega^{2}) A_{0} + \bigl((1-\epsilon^{2})^{2}-
 (1+\epsilon^{2})\omega^{2}\bigr) \frac{A_{0}+B_{0}}{2}   \biggr]. \label{general}
\end{eqnarray}
Then, substituting Eqs.(\ref{a0}) and (\ref{b0}) into Eq.(\ref{general}), we can obtain a result which coincides with
that given in Eq. (\ref{treeres}). Thus, we can reproduce the tree-level decay rate in Eq. (\ref{treeres}) from the 
factorization formula (\ref{fact}). Since we focus on the threshold region of the decay $b \rightarrow c \bar{c} s$
in the present paper, we obtain the threshold expansion of the tree-level decay rate in Eq. (\ref{treeres}) as
\begin{eqnarray}
\Gamma (b \rightarrow c \bar{c} s )_{0}
=\Gamma_{\mu} \frac{4096}{35}\delta^{7/2}\biggl( 1- \frac{231}{66}\delta
 +\frac{1283}{264}\delta^{2}\cdots \biggl),  \label{treeex}
\end{eqnarray}
where the expansion parameter $\delta$ is defined as $\delta = 1/2-\epsilon$.

\begin{figure}[tb]
\begin{center}
\includegraphics[width=13cm]{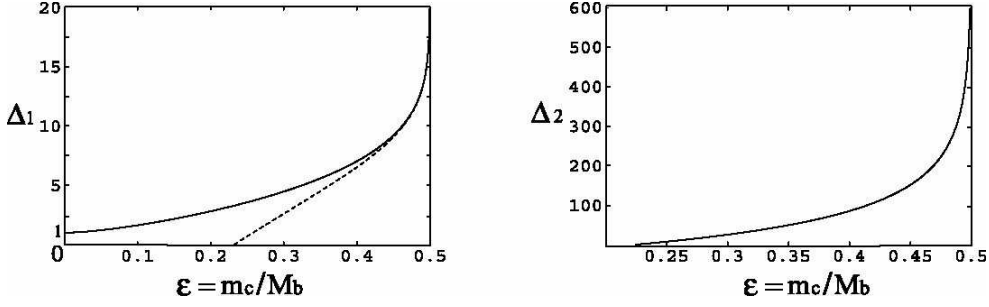}
\end{center} 
\caption{
The corrections $\Delta_{1}$ and $\Delta_{2}$. In the plot of $\Delta_{1}$, the solid curve and the dashed
curve represent the exact results and the first three terms in Eq. (\ref{1three}) in the threshold expansion,
respectively. In the plot of $\Delta_{2}$, the solid curve represents the first three terms in Eq. (\ref{delta2ex}).
 \label{fig1} }
\end{figure}

Next, we consider the estimation of the QCD corrections to the decay $b \rightarrow c \bar{c} s$. In the present paper, we
concentrate on the QCD corrections to the part of the decay  $W^{-} \rightarrow \bar{c} s$ in the factorization
formula (\ref{fact}).\footnote{It is known that penguin contributions to the process $b \rightarrow c \bar{c} s$ exist and that
they are small \cite{alt}. Here we ignore them for simplicity. } 
We write the decay rate including the quantum corrections to the $\bar{c}-s$ current as 
\begin{eqnarray}
\Gamma (b \rightarrow c \bar{c} s ) = \Gamma_{0} \biggl(1+\frac{\alpha_{s}}{\pi}\Delta_{1}
+\biggl(\frac{\alpha_{s}}{\pi}\biggr)^{2} \Delta_{2}\biggr), \label{correction}
\end{eqnarray}
where $\Delta_{1}$ and $\Delta_{2}$ are the ${\cal O}(\alpha_{s})$ and ${\cal O}(\alpha_{s}^{2})$ 
corrections. We write the ${\cal O}(\alpha_{s})$ and ${\cal O}(\alpha_{s}^{2})$ corrections to the
transversal part $A_{0}$ in Eq. (\ref{y}) as  $A_{1}$ and $A_{2}$ and denote the sum of the transversal part as 
$A=A_{0}+A_{1}+A_{2}$. In the same way we write the longitudinal part as
$B=B_{0}+B_{1}+B_{2}$. The  ${\cal O}(\alpha_{s})$ corrections $A_{1}$ and $B_{1}$ are given  
as the functions of $z=\omega^{2}/\epsilon^{2}$ as \cite{vol1}
\begin{eqnarray}
A_{1} &=& \biggl(\frac{\alpha_{s}}{\pi}\biggr) C_{F} A_{0}
\biggl[f_{1}(z)+\frac{2z}{2z+1}f_{2}(z) \biggr], \label{a1} \\
B_{1} &=& \biggl(\frac{\alpha_{s}}{\pi}\biggr) C_{F} B_{0} [f_{1}(z)-1],  \label{b1}
\end{eqnarray}
where $C_{F}=4/3$ and $f_{1}(z)$ and $f_{2}(z)$ are given by
\begin{eqnarray}
f_{1}(z) &=& \frac{13}{4}+2\mbox{Li } \biggl(\frac{1}{z} \biggr)+\mbox{ln} z \ 
 \mbox{ln} \frac{z}{z-1}
-\frac{3}{2}\ \mbox{ln}(z-1)+\mbox{ln}\frac{z}{z-1}+\frac{1}{z} \ \mbox{ln}(z-1)
\nonumber \\ 
&&   +\frac{1}{z-1} \ \mbox{ln} z,  \\
f_{2}(z) &=&  -\frac{5}{2}-\frac{1}{z}-\frac{1}{z-1}+\biggl(\frac{z-1}{z}\biggr) \ 
\biggl(\frac{3}{2}+\frac{1}{2z}\biggr) \ 
\mbox{ln}(z-1) +\frac{z}{(z-1)^{2}} \ \mbox{ln} z.
\end{eqnarray}
Here, $\mbox{Li}(x)=-\int_{0}^{x}\mbox{ln}(1-t)dt/t$ is the standard dilogarithm function. Replacing $A_{0}$
and $B_{0}$ with $A_{1}$ and $B_{1}$, respectively, in Eq. (\ref{general}), we evaluated the integrals numerically
and obtained the form of $\Delta_{1}$ plotted as the solid curve in Fig. \ref{fig1}.
From this graph, we can read off the value $\Delta_{1}=4.46$ at the reference point $\epsilon=0.3$, 
which produces the corrections
$(\alpha_{s}/\pi) \Delta_{1}=0.28$ when we use the value $\alpha_{s}=0.2$. This enhancement by about 30\%
is also reported in Ref.\cite{vol1} . In order to observe the ${\cal O}(\alpha_{s})$ corrections in the threshold region, we use 
the expansion parameter $\rho=1-1/z$ and expand $A_{1}$ and $B_{1}$ in Eqs. (\ref{a1}) and (\ref{b1}) up to the
first three terms as
\begin{eqnarray}
A_{1} &=& \biggl(\frac{\alpha_{s}}{\pi}\biggr) C_{F} \frac{\rho^{2}}{2} \biggl[
\biggl(\frac{9}{2}-3 L_{\rho} +\frac{2}{3}\pi^{2}\biggr) 
+\rho \biggl(-\frac{139}{18}+\frac{11}{3}L_{\rho} -\frac{2}{9}\pi^{2}\biggr) 
\nonumber \\
& & \hspace{70mm} +\rho^{2}\biggl(-\frac{23}{18}+\frac{1}{3}L_{\rho} \biggr) \biggr],  \label{a1ex} \\
B_{1} &=&  \biggl(\frac{\alpha_{s}}{\pi}\biggr) C_{F} (1-\rho) \rho^{2} \biggl[
\biggl(\frac{13}{4}-\frac{3}{2}L_{\rho} +\frac{1}{3}\pi^{2}\biggr)
 -3\rho+\rho^{2}\biggl(-\frac{35}{12}+\frac{1}{2}L_{\rho} \biggr) \biggr].  \nonumber \\
 \label{b1ex}
\end{eqnarray}
These expansions coincide with the results of the perturbative calculations presented in Ref.\cite{cza2} . 
Here, we define $L_{\rho}=\mbox{ln} \rho$.
We change the variable of integration from $\omega^{2}$ to $u=(1/2-\omega)$  in Eq. (\ref{general}) to
obtain the threshold expansion. With $A_{1}$ and $B_{1}$ in Eqs. (\ref{a1ex}) and (\ref{b1ex}), we obtain 
the following threshold expansion :
\begin{eqnarray}
\Gamma_{0} \biggl(\frac{\alpha_{s}}{\pi}\biggl) \Delta_{1}  &=&
6\Gamma_{\mu} \int_{-\delta}^{\delta} du (1-2u) \ \sqrt{u+\delta}
\sqrt{(1-\delta+u)(1+\delta-u)(2-\delta-u)} \nonumber \\
&& \biggl[\Bigl(\frac{1}{2}-u \Bigr)^{2} (1-\delta+u+\delta^{2}-u^{2}) A_{1}  + \nonumber \\
&& \biggl(
\Bigl(\frac{3}{4}+\delta-\delta^{2} \Bigr)^{2}- \Bigl(\frac{1}{2}-u \Bigr)^{2}
\Bigl(\frac{5}{4}-\delta+\delta^{2} \Bigr)
\biggr) \frac{A_{1}+B_{1}}{2}   \biggr]  \\
&=& \Gamma_{\mu} \ \biggl( \frac{\alpha_{s}}{\pi} \biggr) \ \biggl[
\biggl( \frac{387008}{1155}+\frac{16384}{315}\pi^{2}-\frac{16384}{35}\mbox{ln} 2
-\frac{8192}{35} L_{\delta}
\biggr)\delta^{7/2}+  \nonumber \\
&& \biggl( -\frac{125825888}{45045}-\frac{8192}{45}\pi^{2}+\frac{1040384}{315}
\mbox{ln} 2
+\frac{520192}{315} L_{\delta}
\biggr)\delta^{9/2}+  \nonumber \\
&&
\biggl( \frac{22449032}{15015}+\frac{2627584}{10395}\pi^{2}-\frac{727040}{99}
\mbox{ln} 2-\frac{363520}{99} L_{\delta}
\biggr)\delta^{11/2}  \nonumber \\
&& \hspace{70mm} +\cdots  \biggr]. \label{delta1ex}
 \end{eqnarray}
Here, we define $L_{\delta}=\mbox{ln} \delta$. Dividing the quantity in Eq. (\ref{delta1ex}) by that in 
Eq. (\ref{treeex}), we obtain the threshold expansion for $\Delta_{1}$ as
\begin{eqnarray}
\Delta_{1}&=&\biggl( \frac{6047}{2112}+\frac{4}{9}\pi^{2}-4 \ \mbox{ln} 2
-2 L_{\delta} \biggr)+
\biggl( -\frac{570307}{41184}+\frac{128}{9} \ \mbox{ln} 2+\frac{64}{9}L_{\delta}
\biggr)\delta  \nonumber \\
&&+
\biggl( -\frac{11236399}{226512}+\frac{640}{99} \ \mbox{ln} 2+\frac{320}{99} L_{\delta}
\biggr)\delta^{2} +\cdots. \label{1three} 
 \end{eqnarray}
We plot the first three terms in Eq. (\ref{1three}) as the dashed curve in Fig. \ref{fig1}. We can see in the plot 
that although the first three terms in Eq. (\ref{1three}) are sufficient to reproduce the approximate value in the region 
$\epsilon \gtrsim 0.45$, they are not sufficient at the realistic reference point $\epsilon=0.3$.
 
Finally, we obtain the threshold expansion of the ${\cal O}(\alpha_{s}^{2})$ corrections. Their analytic forms,
 $A_{2}$ and $B_{2}$, are not known. The first three terms in the threshold expansions of $A_{2}$
and $B_{2}$ are obtained  as \cite{cza2}
\begin{eqnarray}
A_{2} &=& \biggl(\frac{\alpha_{s}}{\pi}\biggr)^{2} C_{F} \frac{2}{3}\rho^{2}(
C_{F}\Delta_{A}^{v} +C_{A}\Delta_{NA}^{v} +
T_{R}N_{L}\Delta_{L}^{v} +T_{R}\Delta_{H}^{v} ),  \label{a2ex} \\
B_{2} &=& \biggl(\frac{\alpha_{s}}{\pi}\biggr)^{2} C_{F} (1-\rho)\rho^{2}(
C_{F}\Delta_{A}^{s} +C_{A}\Delta_{NA}^{s} +
T_{R}N_{L}\Delta_{L}^{s} +T_{R}\Delta_{H}^{s} ),  \label{b2ex}
\end{eqnarray}
where the explicit forms of $\Delta_{A}^{v}$, $\Delta_{NA}^{v}$, and so on are given in Ref.\cite{cza2} .
Here, we define the coefficients as $C_{F}=4/3, \ C_{A}=3$ and $T_{R}=1/2,$ and we count the number of light
quarks  as $N_{L}=3$ in the present $b$ quark decay. Using Eqs. (\ref{general}), (\ref{treeex}), (\ref{a2ex}) and
(\ref{b2ex}), we obtain the threshold expansion for $\Delta_{2}$ as
\begin{eqnarray}
\Delta_{2}&=&\biggl[ 
\frac{3367549}{76032}+\frac{42877}{12672}\pi^{2}
-\frac{2}{81}\pi^{4}-\frac{444217}{6336}\mbox{ln} 2 -\frac{188}{27} \pi^{2} \mbox{ln} 2
+26 (\mbox{ln} 2)^{2} \nonumber \\ 
&& \ -\frac{444217}{12672}L_{\delta}-\frac{91}{27} \pi^{2} L_{\delta}
+26 \mbox{ln} 2 \ L_{\delta}  +\frac{13}{2} L_{\delta}^{2}-\frac{70}{3}\zeta(3) 
\biggr]+  \nonumber \\
&& \biggl[
-\frac{1471797053}{8895744}+\frac{41175}{9152}\pi^{2}
+\frac{7578347}{28512}\mbox{ln} 2+\frac{464}{243} \pi^{2} \mbox{ln} 2
-\frac{1088}{9} (\mbox{ln} 2)^{2}+ \nonumber \\ 
&& \ \ \frac{7578347}{57024}L_{\delta}+\frac{160}{243} \pi^{2} L_{\delta}
-\frac{1088}{9} \mbox{ln} 2 L_{\delta} -\frac{272}{9} L_{\delta}^{2}+\frac{104}{27}
\zeta(3) \biggr] \delta + \nonumber \\
&& \biggl[ -\frac{29171145503}{3669494400}-\frac{202553411}{12231648}\pi^{2}
+\frac{10217309}{141570}\mbox{ln} 2+\frac{5552}{891} \pi^{2} \mbox{ln} 2 + 
\nonumber \\
&& \ \ \frac{60608}{891} (\mbox{ln} 2)^{2} +\frac{10217309}{283140} L_{\delta}
+\frac{2656}{891} \pi^{2} L_{\delta}+
\frac{60608}{891} \mbox{ln} 2 \  L_{\delta} +
 \nonumber \\
&& \ \frac{15152}{891} L_{\delta}^{2}+\frac{200}{33}\zeta(3) \biggr] \delta^{2} + \cdots.
\label{delta2ex}
\end{eqnarray}
We plot the first three terms in Eq.(\ref{delta2ex}) in Fig. \ref{fig1}.  We can see in the plot that the
divergences in the threshold region are more rapid than the divergences of $\Delta_{1}$ in Eq.(\ref{1three}),
due to the existence of the  terms $L_{\delta}^{2}$. Although the corrections in Eq. (\ref{delta2ex}) at the
reference point $\epsilon=0.3$ are not valid, we read it as $\Delta_{2}=25.2$ for order estimation.
Using this value, the decay rate is estimated as $\Gamma (b \rightarrow c \bar{c} s ) = \Gamma_{0}(1+0.28+0.10)$,
where the second and third terms are the $O(\alpha_{s})$ and $O(\alpha_{s}^{2})$ corrections in Eq. (\ref{correction})
with $\alpha_{s}=0.2$. Then, in order to make the magnitudes of the corrections in the threshold region 
concrete, we take the values $\Delta_{1}=13.3$ and $\Delta_{2}=324$ at the reference point $\epsilon=0.49$.
We estimate the corrections at the reference point as $\Gamma (b \rightarrow c \bar{c} s ) = \Gamma_{0}(1+0.85+1.31)$.
This estimation shows that the sum of the first few terms in the perturbation series at the threshold cannot produce a valid 
approximation.

\section{Coulomb force correction  \label{three}}

\begin{figure}[tb]
\begin{center}
\includegraphics[width=10cm]{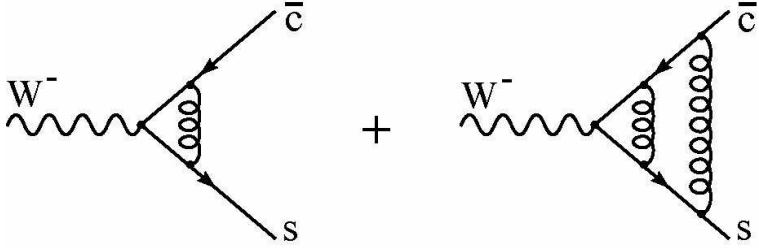}
\end{center} 
\caption{
The Abelian-type diagrams which include the Coulomb-like gluon exchanges
between the anti-charm and strange quarks.   \label{fig2} }
\end{figure}

In this section we obtain the corrections which are produced by the Coulomb force between the 
anti-charm and strange quarks in the final state of the decay $b \rightarrow c \bar{c} s$. Our 
method to obtain the Coulomb force corrections is to replace the plane wave function of the 
strange quark by the wave function in the presence of the Coulomb potential. This method is used to incorporate the
effect of the Coulomb distortion of the electron plane wave by the electromagnetic charge of 
the proton in the neutron beta decay rate \cite{kon}. The decay rate formula in Eq. (\ref{direct})
is not suitable for the distorted wave function by the Coulomb potential because the distorted
wave function is not a momentum eigenfunction and we cannot extract the momentum conservation 
law which is expressed as the delta functions in Eq. (\ref{direct}).\footnote{If we can construct a wave 
packet which coincides with the Coulomb-distorted wave function near
the origin and a plane wave far from the origin, we may be able to include the Coulomb force 
corrections in a more precise way. But in this paper, we do not try to construct such wave packet for 
simplicity.}
We choose a normalization of one particle per volume $V$ for the plane waves in this section. 
Then we start the inclusion of the Coulomb corrections with the decay rate formula as
\begin{eqnarray}
d\Gamma =2\pi |\langle f | \hat{H}_{\mbox{int}} (t=0)  | i \rangle|^{2} 
\ \delta \biggl(M- \sum_{i=1}^{n} p_{i}^{0} \biggr) \prod_{i=1}^{n} \frac{d^{3}p_{i}}
{(2\pi)^{3}} \mbox{V}.  \label{decay2}
\end{eqnarray}
Here, the interaction Hamiltonian is defined as $\hat{H}_{\mbox{int}} (t)=\int d^{3}x 
\hat{{\cal H}}_{\mbox{int}}(\vec{x},t)$, where $\hat{{\cal H}}_{\mbox{int}}$ is interaction Hamiltonian density
and $M$ is the mass of the initial particle. 
Since the initial and final states of the decays are energy eigenstates, we can factor out the 
delta function which gives the energy conservation law. Using the interaction Lagrangian in Eq. (\ref{lag})
for the present decay, we can write the transition matrix as
\begin{eqnarray}
\langle c \bar{c} s | \hat{H}_{\mbox{int}} (t=0)  | b \rangle =
V_{cb}V_{cs}^{*}\frac{G_{F}}{\sqrt{2}}  \int d^{3}x \bigr[ \bar{\psi_{c}} \gamma^{\mu}
(1-\gamma_{5}) \psi_{b} \bigl] \bigr[ \bar{\psi_{s}} \gamma_{\mu}(1-\gamma_{5})
\psi_{\bar{c}} \bigl]. \nonumber \\
 \label{tran}
\end{eqnarray}
In the ordinal perturbation theory, we use the plane waves for the participating particles as
\begin{eqnarray}
\psi_{b}^{r}(\vec{x})&=&\sqrt{\frac{1}{V}} u^{r} (\vec{p}_{b}=0),  \ \ 
\psi_{c}^{r'}(\vec{x}) = \sqrt{\frac{E_{c}+m_{c}}{2E_{c}V}}
 u^{r'} (\vec{p}_{c})e^{+i \vec{p}_{c} \cdot \vec{x}}, \nonumber \\
\psi_{\bar{c}}^{s}(\vec{x}) &=& \sqrt{\frac{E_{\bar{c}}+m_{c}}{2E_{\bar{c}}V}} v^{s} (\vec{p}_{\bar{c}})
e^{-i \vec{p}_{\bar{c}} \cdot \vec{x}}   \ \ \mbox{and} \ \ 
\psi_{s}^{s'}(\vec{x}) = \sqrt{\frac{1}{2V}} u^{s'} (\vec{p}_{s}) e^{+i \vec{p}_{s} \cdot \vec{x}},   \label{plane}
\end{eqnarray}
where the spinors of the particle and anti-particle are defined as
\begin{eqnarray}
u^{s} (\vec{p}) = 
\left(\begin{array}{c}
\chi^{s}  \\ 
\frac{\vec{\sigma}\cdot \vec{p}}{E+m}\chi^{s}
\end{array}\right) \ \ \mbox{and} \ \ 
v^{s} (\vec{p}) = (-s)
\left(\begin{array}{c}
 \frac{\vec{\sigma}\cdot \vec{p}}{E+m}\chi^{-s} \\ 
 \chi^{-s}
\end{array}\right), 
\end{eqnarray}
with $s=1$ for the up spin  and $s=-1$ for the down spin. Here we employ the standard Dirac representation,
in which $\gamma^{0}$ is a diagonal matrix and the plane waves are normalized to one particle per volume
$V$ as $\int_{V} d^{3}x \ \psi^{s}(x)^{\dagger} \psi^{s'}(x)=\delta_{s,s'}$. In order to obtain the Coulomb
corrections, we have the following two approximations. The first approximation is to ignore the small Pauli 
components $(\vec{\sigma}\cdot \vec{p})\chi^{s}/(E+m)$ in the wave functions of the nonrelativistic 
charm and anti-cham quarks, because their small components do not contribute to the leading term of the 
threshold expansion. The second one is to replace the wave function of the strange quark in Eq. (\ref{plane})
with its value at the origin as
\begin{eqnarray}
\psi_{s}^{s'}(\vec{x}) \simeq \psi_{s}^{s'}(\vec{x}=0),  \label{rep1}
\end{eqnarray} 
because the contributions of the wave function at a radius far from the origin can be ignored in the first
term of the threshold expansion. These two approximations are justified by the fact that the resultant decay
rate coincides with the first term in the exact decay rate at tree level in Eq. (\ref{treeex}), as shown below.
Taking advantage of the approximation in Eq. (\ref{rep1}), we replace the plane wave at the origin with the
wave function in the Coulomb potential at the origin. We briefly review the solution of Dirac equation with the Coulomb 
potential (see, e.g, the textbook Ref. \cite{ynd} ). Here we keep the finite mass of the strange quark unless stated otherwise. 
We solve the Dirac equation, 
\begin{eqnarray}
\biggl( i \gamma_{\mu} \partial^{\mu}-|e|Q A_{\mu}\gamma^{\mu}
-m \biggr) \psi(x) =0,    
\end{eqnarray}
with the Coulomb potential $A_{0}=-|e|Q_{s}/r$ and $A_{i}=0$. To solve this equation we can use 
the general form of the wave function in a spherically symmetric potential, 
\begin{eqnarray}
\Psi_{E, +}^{j,j_{3}} (\vec{x},t)=\left(\begin{array}{c}
f(r) {\cal Y}_{j, \ l}^{j_{3}, (+)}   \\ 
g(r) {\cal Y}_{j, \ l+1}^{j_{3}, (-)}, 
\end{array}\right)e^{-iEt},   \label{spher}
\end{eqnarray}
where ${\cal Y}_{j, \ l}^{j_{3}, (\pm)}$ represents the spinor-spherical harmonics, and the total angular momentum
is defined as $j=l+1/2$, with the orbital angular momentum $l$. Although we have one more choice for the
wave function, in which the orbital angular momentum of the upper two components is larger than that of 
the lower two components by unity, we use the wave function in Eq. (\ref{spher}), because the s-wave of
the particle (strange quark) in the present case contributes to the Coulomb corrections. We obtain the
solutions of the radial parts with the continuum energy spectrum as  
\begin{eqnarray}
&&\left(\begin{array}{r}
f_{l} (r)    \\ 
g_{l} (r)
\end{array}\right) =
N_{l} e^{-ikr}(2kr)^{\gamma-1}  \times \nonumber \\
&& \hspace{5mm} \left(\begin{array}{c} 
\sqrt{E+m} \bigl[ {}_{1}F_{1}(\xi,1+2\gamma,2ikr) + \frac{\xi}{l+1-ia} 
{}_{1}F_{1}(\xi+1,1+2\gamma,2ikr) \bigr]  \\ 
\sqrt{E-m} \bigl[ {}_{1}F_{1}(\xi,1+2\gamma,2ikr) - \frac{\xi}{l+1-ia} 
{}_{1}F_{1}(\xi+1,1+2\gamma,2ikr) \bigr]
\end{array}\right),  \label{radial}
\end{eqnarray}
where ${}_{1}F_{1}$ is the Kummer function and we have $\gamma=\sqrt{(l+1)^{2}-\alpha_{0}^{2}},
\ \xi=\gamma + i \nu, \ \nu=\alpha_{0}E/k,  \ a=\alpha_{0}m/k$ and $\alpha_{0}=-|e|^{2}QQ_{s}$.
Using the normalization condition
\begin{eqnarray}
\int d^{3} x\  \Psi_{E}^{j,j_{3}} (\vec{x})^{\dagger} \Psi_{E}^{j',j_{3}'} (\vec{x}) = \delta_{j,j'}
\delta_{j_{3},j_{3}'}\frac{E\pi}{k^{2}}\delta(k-k'),
\end{eqnarray}
we fix the normalization constant as 
\begin{eqnarray}
N_{l}=e^{\pi \nu/2} \cdot  \frac{|\Gamma(1+\gamma+i\nu)|}
{|\Gamma(1+2\gamma)|},  \label{norm}
\end{eqnarray}
where $\Gamma(z)$ is the Gamma function. We can rewrite the solutions of the radial parts in Eq. (\ref{radial})
near the origin as
\begin{eqnarray}
\left(\begin{array}{r}
f_{l}(r=R)    \\ 
g_{l}(r=R)
\end{array}\right) \simeq 
N_{l} e^{-ikr}(2kR)^{\gamma-1} 
\left(\begin{array}{c}
\sqrt{E+m} \bigl[ 1+\frac{\xi}{1-ia}  \bigr]  \\ 
\sqrt{E-m} \bigl[1- \frac{\xi}{1-ia} \bigr]
\end{array}\right),   \label{close}
\end{eqnarray}
where we define the radius near the origin as $R\simeq 0$. When the Coulomb potential is switched off and
the Coulomb wave is reduced to a free spherical wave, the inner product of the wave functions in Eq. (\ref{spher}) 
with $j=1/2$ near the origin is given by
\begin{eqnarray}
\Psi_{E}^{j=1/2, j_{3}} (\vec{x})^{\dagger} \Psi_{E}^{j=1/2, j_{3}} (\vec{x})
\bigl|_{r =R, \ \alpha_{0}=0}= \frac{E+m}{4\pi},
\end{eqnarray}
which differs from that of the plane waves, $\psi^{s}(x)^{\dagger} \psi^{s}(x) =1/V$. In order to replace 
the plane wave in Eq. (\ref{rep1}) with the Coulomb-distorted wave in Eq. (\ref{spher}) near the origin in a 
consistent way, we modify the normalization constant of the Coulomb wave in Eq. (\ref{norm}) as 
\begin{eqnarray}
\Psi_{E}^{j,j_{3}} (\vec{x})^{\mbox{new}}\bigl|_{r =R} =\sqrt{\frac{4\pi}{(E+m)V}} \Psi_{E}^{j,j_{3}}
 (\vec{x}) \bigl|_{r =R}.  \label{new}
\end{eqnarray}
For this purpose, here we define the Fermi function as 
\begin{eqnarray}
\mbox{F}(E, \alpha_{0})=\frac{|\Psi_{E}^{j,j_{3}} (\vec{x},\alpha_{0})|^{2}}
{|\Psi_{E}^{j,j_{3}} (\vec{x},\alpha_{0}=0)|^{2}} \Biggl|_{r=R},
\end{eqnarray}
with $R \simeq 0$. Using Eqs. (\ref{spher}), (\ref{norm}), and (\ref{close}), we obtain the explicit form of the
Fermi function as
\begin{eqnarray}
\mbox{F}(E, \alpha_{0})=4\frac{E+m \gamma_{0}}{E+m } \cdot (2kR)^{2(\gamma_{0}-1)}
\cdot e^{\pi \nu}
\cdot  \frac{|\Gamma(\gamma_{0}+i\nu)|^{2}}{|\Gamma(1+2\gamma_{0})|^{2}},  \label{fermifun}
\end{eqnarray}
with the relations $E=\sqrt{m^{2}+k^{2}}$ and $\gamma_{0}=\sqrt{1-\alpha_{0}^{2}}$. Hereafter, we omit
the mass of the strange quark again and write the energy of the strange quark as $E_{s}=|\vec{p}_{s}|$. 
Substituting the plane waves in Eq. (\ref{plane}) into Eq. (\ref{tran}), except for the strange quark, and using 
the renormalized Coulomb wave in Eq. (\ref{new}) for the strange quark, we obtain the transition matrix element as
\begin{eqnarray}
\langle f | \hat{H}_{\mbox{int}} (t=0)  | i \rangle &=& V_{cb}V_{cs}^{*} \frac{\sqrt{2}G_{F}}{V} (2\pi)^{3} 
\delta^{3}(\vec{p}_{c}+ \vec{p}_{\bar{c}}) \bigl[ \delta_{r,r'} j^{0}(\bar{c}s)   \nonumber \\
&& \hspace{40mm}  +  (\chi^{r' \dagger} \sigma^{i}
 \chi^{r}) j^{i}(\bar{c}s) \bigr]  \label{fgt} \\
&=& \langle f | \hat{H}_{\mbox{int}} | i \rangle_{V} +\langle f | \hat{H}_{\mbox{int}}| i \rangle_{A}, \label{tran2}
\end{eqnarray}
where $\langle f | \hat{H}_{\mbox{int}} | i \rangle_{V}$ and $\langle f | \hat{H}_{\mbox{int}}| i \rangle_{A}$
are defined as the first and second terms in Eq. (\ref{fgt}), referring to the vector and the axial vector parts of 
the bottom-charm
current, respectively. We write the anti-charm and strange quark current as $j^{\mu}(\bar{c} s) =\bar{\Psi}_{s} 
\gamma_{\mu}L \psi_{\bar{c}}= \bar{\Phi}_{s} \gamma_{\mu} \phi_{\bar{c}}$, where the left-handed fields are
written as $\phi_{\bar{c}}=L \psi_{\bar{c}}$ and $\Phi^{s}=L  \Psi_{E}^{j, j_{3} \mbox{new}}$, with the projection operator
$L=(1-\gamma_{5})/2$. We average the square of Eq. (\ref{tran2}) over the spins and obtain the contributions of
the vector and axial vector parts as
\begin{eqnarray}
\frac{1}{2} \sum_{r,r',s,s'} |\langle f | \hat{H}_{\mbox{int}} | i \rangle_{V} |^{2} &=& |V_{cb}V_{cs}|^{2} 
\frac{G_{F}^{2}}{V^{2}}
\frac{(2\pi )^{3} \delta^{3}(\vec{p}_{c}+ \vec{p}_{\bar{c}})}{V} \ \mbox{F}(E_{s},\alpha_{0}),  \label{square1} \\
\frac{1}{2}  \sum_{r,r',s,s'} |\langle f | \hat{H}_{\mbox{int}}| i \rangle_{A} |^{2} &=& |V_{cb}V_{cs}|^{2} 
 \frac{G_{F}^{2}}{V^{2}}
\frac{(2\pi)^{3} \delta^{3}(\vec{p}_{c}+ \vec{p}_{\bar{c}})}{V} \ 3 \ \mbox{F}
(E_{s},\alpha_{0}),   \label{square2}
\end{eqnarray}
where the interference term of the vector and axial vector parts does not contribute to the spin-averaged decay rate. 
When we derive Eqs. (\ref{square1}) and (\ref{square2}), we use the formulae $\sum_{s} \phi_{\bar{c}}^{s}(x) 
\phi_{\bar{c}}^{s}(x)^{\dagger}\simeq L/2V$ and $\sum_{s} \Phi^{s} (\vec{x})^{\dagger} \Phi^{s}(\vec{x})
\bigl|_{r =R}=F(E_{s}, \alpha_{0})/V$. Substituting Eqs. (\ref{square1}) and (\ref{square2}) into Eq. (\ref{decay2}), 
we obtain the decay rate including the Fermi function $\Gamma(b \rightarrow  c \bar{c} s)^{c}$  as
\begin{eqnarray}
\Gamma(b \rightarrow  c \bar{c} s)^{c}
= 768 \ \Gamma_{\mu} \int_{0}^{\delta} dz z^{2}
(1 - 2z)  \sqrt{\delta -z -\delta^{2} +z^{2}} \cdot
 \mbox{F}(E_{s},\alpha_{0}), 
\end{eqnarray}
where the variable of integration is defined as $z=E_{s}/2M_{b}$. We compute only the contribution
to first order in $\delta$ and obtain the decay rate as
\begin{eqnarray}
\Gamma (b \rightarrow  c \bar{c} s)^{c}
= \Gamma_{0}^{c} \biggl[  1+ \pi \alpha_{0} +\biggl(
\frac{1019}{210}-\gamma_{e}+\frac{\pi^{2}}{3} -\mbox{ln}\bigl(M_{b} R \delta \bigr)
 -4 \mbox{ln}2 \biggr) \alpha_{0}^{2} \biggr]. \nonumber \\
 \label{series}
\end{eqnarray}
Here, we have used the expansion of the Fermi function in the coupling constant, $\mbox{F}(E, \alpha_{0}) = 1+
\pi \alpha_{0} +(\pi^{2}/3+3-\gamma_{e} -\mbox{ln}(2ER) ) \alpha_{0}^{2} + \cdots$, and written the 
tree-level rate as $\Gamma_{0}^{c}=\Gamma_{\mu} (4096/35)\delta^{7/2}$, which coincides with the 
first term in Eq. (\ref{treeex}). Here, $\gamma_{e}\simeq 0.577$ is Euler's constant. Since the Coulomb corrections 
originate in the effects of the Coulomb-like gluon exchanges between the $\bar{c}$ and $s$ quarks,
we can expect that the ${\cal O}(\alpha_{0})$ and ${\cal O}(\alpha_{0}^{2})$ corrections in Eq. (\ref{series})
are included in the Abelian-type one- and two-loop diagrams shown in Fig. \ref{fig2}.
In the correspondence, we can introduce the strong coupling constant as $\alpha_{0}=C_{F} \alpha_{s}$ and 
rewrite the series in Eq. (\ref{series}) as 
\begin{eqnarray}
\frac{\Gamma (b \rightarrow  c \bar{c} s)^{c}}{\Gamma_{0}^{c}}
&=& 1+  C_{F}\pi^{2} \biggr(\frac{\alpha_{s}}{\pi}\biggl)
+\biggl( \frac{1019}{210}-\gamma_{e}+\frac{\pi^{2}}{3} -\mbox{ln}\bigl(M_{b} R \delta \bigr)
 -4 \mbox{ln}2 \biggr)  \nonumber \\
&& \hspace{60mm} \times \ C_{F}^{2}\pi^{2} \biggr(\frac{\alpha_{s}}{\pi}\biggl)^{2}.   \label{series2}
\end{eqnarray}
Here we compare the corrections in Eq. (\ref{series2}) with the abelian part of the perturbative corrections 
in Eqs. (\ref{1three}) and (\ref{delta2ex}).
In the comparison we assume that the correspondence of terms is categorized according to
the powers of the quantities, $\pi$, $\zeta(3)$, and logarithmic function. For example, the  $\pi^{2}L_{\delta}$
term at  ${\cal O}(\alpha_{s}^{2})$ in Eq. (\ref{series2}) is compared with the $\pi^{2}L_{\delta}$
term in the abelian part of the perturbative corrections in Eq. (\ref{delta2ex}),
rather than with the $L_{\delta}$ term.
The coefficient $C_{F}\pi^{2}$ in the ${\cal O}(\alpha_{s})$ correction is three times larger than the similar 
$\pi^{2}$ term, $4\pi^{2}/9$, in Eq. (\ref{1three}). We conjecture that 
this difference is compensated for other corrections which have the other physical origins. At ${\cal O}(\alpha_{s}^{2})$,
we extract the Abelian part from $\Delta_{2}$ in Eq. (\ref{delta2ex}) 
by taking only $\Delta_{A}^{v}$ and $\Delta_{A}^{s}$ in Eqs. (\ref{a2ex}) and (\ref{b2ex}) and thereby obtain 
the first term as
\begin{eqnarray}
\Delta_{2}^{\mbox{abel}}&=&  C_{F}^{2} \biggl[ 
\frac{119005}{22528}+\frac{1091}{2816}\pi^{2}
+\frac{1}{90}\pi^{4}-\frac{16381}{2816}\mbox{ln} 2-\frac{2}{3} \pi^{2} \mbox{ln} 2
+\frac{9}{2} (\mbox{ln} 2)^{2} \nonumber \\ 
&&-\frac{16381}{5632}L_{\delta}-\frac{5}{6} \pi^{2} L_{\delta}
+\frac{9}{2} \mbox{ln} 2 \  L_{\delta}  +\frac{9}{8} L_{\delta}^{2}-3 \ \zeta(3) 
\biggr] +\cdots. \label{abel}
\end{eqnarray}
The term $-C_{F}^{2}\pi^{2}L_{\delta}$ in the ${\cal O}(\alpha_{s}^{2})$ corrections in Eq. (\ref{series2})
accounts for 120\% of the similar $\pi^{2}L_{\delta}$ term, $-(5/6)C_{F}^{2}\pi^{2}L_{\delta}$, in Eq. (\ref{abel}).  
Because the $L_{\delta}^{2}$ and $L_{\delta}$ terms are singular at the threshold, these terms
are not cancelled by other non-logarithmic terms. Based on the assumption that the terms are 
categorized according to $L_{\delta}$, $\pi^{2}$, and so on, we conjecture that the sum of the Coulomb corrections
and other $\pi^{2}L_{\delta}$ corrections, which come from other physical origins in the Abelian 
theory, is $-(5/6)C_{F}^{2}\pi^{2}L_{\delta}$ in  Eq. (\ref{abel}). The numerical impact of 
the Coulomb corrections, however, is not clear at the moment. 
The differences among the other terms, $1019\pi^2/210, \ \pi^{4}/3,$ and $-4 \pi^2 \mbox{ln}2$, at
the ${\cal O}(\alpha_{s}^{2})$ in Eq. (\ref{series2}) and the corresponding terms in Eq. (\ref{abel}) are large. 
Although the Coulomb corrections do not contain terms of the forms $\mbox{ln}2, (\mbox{ln}2)^{2},
L_{\delta}, \mbox{ln}2 L_{\delta}, L_{\delta}^{2}, \mbox{and} \  \zeta(3)$, the perturbative corrections
do contain such terms. Further, the Coulomb corrections contain $\gamma_{e}$, while the perturbative corrections
do not. We also believe that these differences are compensated for the other corrections.

\section{Hybrid anomalous dimension   \label{four}}
In this section, we show that the leading and next-to-leading logarithmic terms in the first order of the threshold 
expansions in Eqs. (\ref{1three}) and  (\ref{delta2ex}) originate in the anomalous dimension of the heavy-light 
$\bar{c}-s$ current $j=\bar{q} \Gamma  Q$ in the heavy quark effective field theory (HQET), which is called the
hybrid anomalous dimension \cite{shi}. Here, $q$ is a massless quark field, $Q$ is a heavy quark field in HQET, 
which satisfies the relation $Q=\gamma^{0}Q$, and $\Gamma$ represents the Dirac gamma matrices. First,
solving the renormalization group equation
\begin{eqnarray}
\biggl(\mu \frac{d}{d \mu} -\gamma_{h} (\alpha_{s}(\mu)) \biggr){\cal C}(\mu, M)=0,  \label{ren}
\end{eqnarray}
we obtain the Wilson coefficient for the hybrid anomalous dimension as
\begin{eqnarray}
{\cal C}(\mu, M) = {\cal C}(M, M)\exp \biggl[-\int^{\alpha_{s}^{(n_{l})}(\mu)}_{\alpha_{s}^{(n_{l})}(M)} 
\frac{\gamma_{h} (\alpha_{s})}{2\beta (\alpha_{s})}
\frac{d \alpha_{s}}{\alpha_{s}} \biggr], \label{coe}
\end{eqnarray}
where the hybrid anomalous dimensions in HQET are given in Refs.\cite{shi, pol1, pol2, jix} and \cite {che2} as
\begin{eqnarray}
\gamma_{h} = \gamma_{h}^{(0)} \frac{\alpha_{s}}{4\pi} +\gamma_{h}^{(1)} 
\biggl(\frac{\alpha_{s}}{4\pi}\biggr)^{2}, \label{had} 
\end{eqnarray}
with $\gamma_{h}^{(0)}=-4$ and $\gamma_{h}^{(1)} =-\frac{254}{9}+\frac{20}{9}n_{l}-
\frac{56}{27}\pi^{2}$. The beta function are given by
\begin{eqnarray}
\beta &=& \beta_{0} \frac{\alpha_{s}}{4\pi} +\beta_{1}
\biggl(\frac{\alpha_{s}}{4\pi}\biggr)^{2}, \label{beta}
\end{eqnarray}
with $\beta_{0}=11-(2/3)n_{l}$ and $\beta_{1}=102-(38/3)n_{l}$. The hybrid anomalous dimension
in HQET does not depend on the structure of the gamma matrices $\Gamma$.
We write the initial condition in Eq. (\ref{coe}) as ${\cal C}(M, M)$, 
which depends on the structure $\Gamma$. In order to deal with the universal part in Eq. (\ref{coe}), we set
the initial condition as ${\cal C}(M, M)=1$ hereafter. Valid initial conditions 
are introduced later through the matching coefficients.
Using Eqs. (\ref{had}) and (\ref{beta}), we obtain ${\cal C}(\mu, M)$ in Eq. (\ref{coe}) as 
 \begin{eqnarray}
{\cal C}(E_{s}, m_{c})   =\biggl[ \frac{\alpha_{s}(E_{s})}{\alpha_{s}(m_{c})} \biggr]^{-\frac{\gamma_{h}^{(0)}}
{2\beta_{0}}} \exp \Biggl[ -\frac{\gamma_{h}^{(0)}}{8\pi\beta_{0}}  
\Biggl( \frac{\gamma_{h}^{(1)}}{\gamma_{h}^{(0)}}- \frac{\beta_{1}}{\beta_{0}} \Biggr) 
(\alpha_{s}(E_{s})-\alpha_{s}(m_{c})) \Biggr],  \label{coeex}
\end{eqnarray}
where we write $\mu$ as the threshold momentum $E_{s}$, like the energy of the strange quark in the presently
considered decay, $b \rightarrow c\bar{c} s$, and $M$ as a heavy quark mass $m_{c}$, like the mass of the charm quark. 
Furthermore, we substitute $\alpha_{s}(E_{s})$ to Eq. (\ref{coeex}) with the running coupling constant up to 
two-loop order,
\begin{eqnarray}
\alpha_{s}(E_{s})=\frac{\alpha_{s}(m_{c})}{1+\alpha_{s}(m_{c})
\frac{\beta_{0}}{2\pi}\log\frac{E_{s}}{m_{c}}
+\alpha_{s}(m_{c})^{2}\frac{\beta_{1}}{8\pi^{2}}\log\frac{E_{s}}{m_{c}}},     \label{run}
\end{eqnarray}
and obtain the square of the Wilson coefficient in Eq. (\ref{coeex}) as the following series in powers of the strong 
coupling constant $\alpha_{s}(m_{c})$ :
\begin{eqnarray}
{\cal C}^{2}(E_{s}, m_{c})  &=&
1+\biggl(\frac{\alpha_{s}(m_{c})}{\pi}\biggr) \biggl[\frac{\gamma_{h}^{(0)}}{2} \mbox{ln} \ \frac{E_{s}}
{m_{c}}\biggr]
\nonumber \\
&+&\biggl(\frac{\alpha_{s}(m_{c})}{\pi}\biggr)^{2} \biggl[
\frac{1}{8}\gamma_{h}^{(0)}(\gamma_{h}^{(0)}-\beta_{0})\biggl(\mbox{ln} \ \frac{E_{s}}{m_{c}}\biggr)^{2}
+\frac{\gamma_{h}^{(1)}}{8}\biggl(\mbox{ln} \ \frac{E_{s}}{m_{c}}\biggr)
\biggr]. \label{coes}
\end{eqnarray}
The leading logarithmic terms in the ${\cal O}(\alpha_{s})$ and ${\cal O}(\alpha_{s}^{2})$ corrections
in Eq. (\ref{coes}) with $n_{l}=3$ coincide with those in the perturbative corrections in Eq. (\ref{1three})
and (\ref{delta2ex}) when $\delta$ is replaced with $E_{s}/m_{c}$. This agreement shows that the leading
logarithmic terms in the perturbative corrections come from the hybrid anomalous dimension. 
By contrast, the next-to-leading logarithmic term at ${\cal O}(\alpha_{s}^{2})$ does not coincide with that in 
Eq. (\ref{delta2ex}). This disagreement between the next-to-leading logarithmic terms is natural for the
following two reasons. The first reason is that the next-to-leading logarithms in the transversal part, $A/A_{0}$,
and the longitudinal part, $B/B_{0}$, in the factorization formula (\ref{general}) are different,
while the leading logarithmic terms are the same. The second reason is that although the product of the leading
logarithmic term and the non-logarithmic terms at ${\cal O}(\alpha_{s})$ can produce the next-to-leading
logarithmic terms at ${\cal O}(\alpha_{s}^{2})$, the square of the Wilson coefficient in Eq. (\ref{coes}) 
does not contain non-logarithmic terms at ${\cal O}(\alpha_{s})$. In connection with the first of these reasons, 
the relations between $A/A_{0}$, $B/B_{0}$ and the universal part, $R(\rho)$,  are given by \cite{che}
\begin{eqnarray}
\frac{A(\rho)}{A_{0}(\rho)} &=& C_{v}(m_{c})^{2} R(\rho),  \label{afac} \\  
\frac{B(\rho)}{B_{0}(\rho)} &=& \biggl[\frac{ \bar{m}_{c}(m_{c}) }{m_{c}}C_{s}(m_{c}) 
\biggr]^{2} R(\rho),  \label{bfac}
\end{eqnarray}
where the matching coefficients are obtained as \cite{gro}
\begin{eqnarray}
C_{v}(m_{c}) &=& 1-C_{F}\biggl(\frac{\alpha_{s}(m_{c})}{\pi}\biggr),  \\
C_{s}(m_{c}) &=& 1+\frac{C_{F}}{2}\biggl(\frac{\alpha_{s}(m_{c})}{\pi}\biggr), \\
\frac{\bar{m}_{c}(m_{c})}{m_{c}} &=& 1- \frac{4}{3}\biggl(\frac{\alpha_{s}(m_{c})}
{\pi}\biggr).
\end{eqnarray}
Using Eqs. (\ref{a0}) and (\ref{a1ex}) and the leading and next-to-leading logarithmic terms in (\ref{a2ex}),
we can extract the universal part $R(\rho)$ in Eq. (\ref{afac}) as
\begin{eqnarray}
R(\rho) &=&
1  + \biggl(\frac{\alpha_{s}(m_{c})}{\pi}\biggr) \biggl[-2 L_{\rho} +\frac{17}{3}+
\frac{4}{9}\pi^{2} \biggr]  +\biggl(\frac{\alpha_{s}(m_{c})}{\pi}\biggr)^{2} \nonumber \\
&& \times \biggl[\biggl(\frac{15}{2}-\frac{n_{l}}{3}\biggr) L_{\rho}^{2} 
+\biggl(-\frac{1657}{36}-\frac{97}{27}\pi^{2}
+n_{l}\Bigl(\frac{13}{6}+\frac{4}{27}\pi^{2} \Bigr)
\biggr) L_{\rho}\biggr],   \label{uni}
\end{eqnarray}
which is consistent with Eq. (\ref{bfac}). In connection to the second reason stated above, regarding
the lack of non-logarithmic terms, 
we consider the non-logarithmic terms at ${\cal O}(\alpha_{s})$ in Eq. (\ref{uni}) and introduce the 
following factor at the scale of the threshold momentum $E_{s}$ :
\begin{eqnarray}
D(E_{s}) =1+\biggl(\frac{\alpha_{s}(E_{s})}{\pi}\biggr)  \ \biggl(\frac{17}{3}+\frac{4}{9}\pi^{2}  \biggr). \label{facd}
\end{eqnarray}
The relation between $\alpha_{s}(E_{s})$ and $\alpha_{s}(m_{c})$ appearing here is given in Eq. (\ref{run}).
When the coefficient ${\cal C}^{2}(E_{s}, m_{c})$ is improved by the factor $D(E_{s})$, it coincides with
the universal part in Eq. (\ref{uni}), up to the next-to-leading logarithms at ${\cal O}(\alpha_{s}^{2})$, as 
\footnote{In Ref.\cite{che} , The universal part, $R(\rho)$, is obtained by using Eq. (\ref{ren}) and the ${\cal O}(\alpha_{s})$ 
corrections of $R(\rho)$. In the same way, here we obtained the relation (\ref{cdr}) by introducing 
the factor in Eq. (\ref{facd})  which has the non-logarithmic terms at ${\cal O}(\alpha_{s})$ in $R(\rho)$
and is defined at the energy scale $E_{s}$.}
\begin{eqnarray}
{\cal C}^{2}(E_{s}, m_{c})  \cdot D(E_{s}) = R\biggl(\rho=\frac{E_{s}}{m_{c}}\biggr). \label{cdr}
\end{eqnarray}
This means that the next-to-leading logarithmic terms in  Eq. (\ref{delta2ex}) can be reduced to the hybrid 
anomalous dimension and the running coupling constant in Eq. (\ref{run}).

\section{Summary  \label{five}}
We have studied the physical origins of the corrections to the $\bar{c}-s$ current of the decay $b \rightarrow c\bar{c} s$
at the threshold. We first obtained the ${\cal O}(\alpha_{s})$ and ${\cal O}(\alpha_{s}^{2})$ perturbative
corrections in Eqs. (\ref{1three}) and (\ref{delta2ex}), respectively. Second, we obtained the corrections in 
Eq. (\ref{series2}), which are produced by the Coulomb force between the $\bar{c}$ and $s$ quarks. The Coulomb
corrections, $C_{F}\pi^{2}$ at ${\cal O}(\alpha_{s})$ and $-C_{F}^{2}\pi^{2}\mbox{ln} \ \delta$ at
${\cal O}(\alpha_{s}^{2})$, account for 300\% and 120\% of the terms with the same forms in the Abelian parts of the 
perturbative corrections, respectively. The differences among the other terms in the ${\cal O}(\alpha_{s}^{2})$
corrections in Eq. (\ref{series2}) and the perturbative corrections of the same form are large. 
These differences  should be compensated by other corrections which have other physical origins.
We finally confirmed that the Wilson coefficient for the hybrid anomalous dimension reproduces the leading logarithmic
terms in the perturbative corrections.  We also confirmed that when the Wilson coefficient is improved by the 
non-logarithmic terms in the ${\cal O}(\alpha_{s})$ perturbative corrections,  the improved coefficient can 
produce the next-to-leading logarithms in the ${\cal O}(\alpha_{s}^{2})$ perturbative corrections. We 
have identified three challenging problems in the present paper. The first problem is to reproduce the 
Coulomb corrections in Eq. (\ref{series2}) with perturbative calculations. In such calculation
we should calculate only the loop diagrams shown in 
Fig \ref{fig2} and extract only the soft-gluon contributions in the loop integrals. The second problem is to 
determine the accuracy of the method used to incorporate the Coulomb force corrections in \S \ref{three}. 
Solving this problem, we will be able to identify those terms that come from the Coulomb force corrections 
with the higher precision. The last problem 
is to find the relations between the two next-to-leading logarithmic terms of the Coulomb corrections in Eq. (\ref{series2})
and the Wilson coefficient for the hybrid anomalous dimension  in Eq. (\ref{uni}). We should observe only the Abelian
part of the hybrid anomalous dimension. Although the two next-to-leading logarithmic terms may originate in the soft and 
hard regions of the loop momentum, respectively, these relations are not trivial. We need more sophisticated studies 
to reveal these relations.

\subsection*{Acknowledgment}
I would like to thank A. Pak for the valuable discussions. This work is supported by the fund for
the Science and Engineering Research Canada.

\end{document}